\documentclass[12pt,a4paper]{article}

 \usepackage{a4wide}
 \usepackage{latexsym}
 \usepackage{epsf}
 \usepackage{amssymb}
 \usepackage{amsmath, cite}
 \usepackage{slashed,epsfig}

\begin{document}

\begin{flushright}
\small
UG-04-04\\
{\bf hep-th/0407047}\\
\date \\
\normalsize
\end{flushright}

\begin{center}


\vspace{.7cm}

{\LARGE {\bf Scalar Cosmology with }} \\

\vspace{.7cm}

{\LARGE {\bf Multi-exponential Potentials}}

\vspace{1.2cm}

{\large Andres Collinucci, Mikkel Nielsen and Thomas Van Riet}
\vskip 1truecm

\small
{\it Centre for Theoretical Physics, University of Groningen,\\
   Nijenborgh 4, 9747 AG Groningen, The Netherlands\\
E-mail: {\tt (a.collinucci, m.nielsen, riett)@phys.rug.nl}}

\vspace{.7cm}


{\bf Abstract}

\end{center}

\begin{quotation}

\small

We investigate cosmologies with an arbitrary number of scalars and
the most general multi-exponential potential. By formulating the
equations of motion in terms of autonomous systems we complete the
classification of power-law and de Sitter solutions as critical
points, e.g.~attractor and repeller solutions, in terms of the
scalar couplings. Many of these solutions have been overlooked in
the literature.

We provide specific examples for double and triple exponential
potentials with one and two scalars, where we find numerical
solutions, which interpolate between the critical points. Some of
these correspond to the reduction of new exotic S-brane solutions.

\end{quotation}

\newpage

\pagestyle{plain}

\section{Introduction}

The discovery that the universe might currently be in a phase of
accelerate expansion \cite{Riess:1998cb,Perlmutter:1998np} has led
to a large interest in finding de Sitter solutions or more general
accelerating cosmologies from M-theory, see e.g.~
\cite{Fre:2002pd,deRoo:2002jf,Kachru:2003aw,Townsend:2001ea,Cornalba:2002fi,Townsend:2003fx,
Ohta:2003pu,Roy:2003nd,Chen:2003ij,Wohlfarth:2003kw,Bergshoeff:2003vb,Jarv:2004uk,Ohta:2003ie}
and references therein.

A simple way to study accelerating cosmologies is to consider
models containing just gravity and a number of scalars with a
potential. This has a long history and has resulted in models for
inflation \cite{Linde:1982mu}, describing the early universe, and
later for quintessence \cite{Caldwell:1998je}, where a scalar
field gives rise to a small effective cosmological constant.
Multi-exponential potentials comprise a specific class of
potentials which have been studied a lot, and these are of
interest for two reasons. First, they can arise from M-theory in
many ways, e.g. via compactifications on product spaces possibly
with fluxes
\cite{Ivashchuk:1989gz,Lu:1997jk,Lukas:1997iq,Chen:2003dc}, and
second, the equations of motion can be written as an autonomous
system. This approach allows for an algebraic determination of
power-law and de Sitter solutions, i.e.~critical points, which can
correspond to early- and late-time asymptotics of general
solutions. Many authors have made use of this fact, see e.g.~
\cite{Halliwell:1987ja,Coley:1997nk,Coley:1999mj,Copeland:1998et,Liddle:1998jc,Heard:2002dr,Bergshoeff:2003vb,Guo:2003eu}
and references therein.

The understanding of multi-exponential potentials has gradually
evolved over the years. In the early days the single exponential
was studied in the context of inflation, where it was discovered
that this potential allowed for a power-law solution
\cite{Halliwell:1987ja}. Later on the effect of additional
exponential terms each carrying a different scalar was studied.
This model is called ``assisted inflation'' \cite{Liddle:1998jc}.
The outcome is that the scalars `assist' each other, in the sense
that each term contributes in the same way to the power-law
behaviour of the scale factor and all the contributions are added.
Later on the effect of a cross coupling between scalars was
searched for, this resulted in a model called ``generalized
assisted inflation'' \cite{Copeland:1999cs}. It was shown that
these multi-exponential potentials also allowed for power-law
solutions. However the understanding of multi-exponential
potentials wasn't complete. The class of potentials described in
\cite{Copeland:1999cs} doesn't cover all the possible
multi-exponential potentials. There was a strong restriction on
the scalar couplings in their model, such that it only allows for
power-law solutions, namely the potentials don't have any extrema.
But, a considerable amount of models nowadays which are inspired
by string theory seem to be multi-exponentials with extrema (which
allow for de Sitter solutions). Hence they don't fall in the class
of generalized assisted inflation.

The goal of this paper is to study the most general
multi-exponential potential. This is done using the elegant
formalism of autonomous dynamical systems. We construct all
possible power-law solutions and de Sitter solutions by
constructing the critical points to which they correspond in this
formalism. We point out that even in the case of generalized
assisted inflation many power-law solutions which correspond to so
called non-proper critical points were overlooked. To illustrate
this we consider the special cases of double and triple
exponential potentials with one or two scalars. These can arise
from M-theory, and the interpolating solutions then correspond to
the reduction of S-branes \cite{Gutperle:2002ai} and so-called
exotic S-branes \cite{Bergshoeff:2003vb}, respectively. For the
exotic S-branes we derive the phases of accelerate expansion and
find special cases where the number of such phases can be
arbitrarily high. This can be useful for solving the cosmological
coincidence problem, since oscillating dark energy could give an
explanation of why we see a recent take over of dark energy in our
present universe. It would simply be an event that occurs for many
times during the evolution of the universe.

The paper is organised as follows. In section 2 we present the
system consisting of gravity and scalars with a potential. In
section 3 we perform the general analysis of critical points. In
section 4 we consider the special cases of double exponentials. In
section 5 we present cases which can be obtained from the
reduction over a three-dimensional group manifold. Finally, we end
with a discussion of our results in section 6.

\section{Scalar Gravity with Multi-exponential Potentials}

We consider 4-dimensional spatially flat FLRW gravity with $N$
scalars $\phi_I$ which only depend on (cosmic) time $\tau$. The
scalars have a potential which is of the most general exponential
form:
\begin{equation} \label{potential}
V\vec{(\phi)}=\sum_{i=1}^m
\Lambda_{i}\,e^{-\vec{\alpha_{i}}\cdot\vec{\phi}}\,.
\end{equation}
Thus, the scalar potential is characterised by $m$ vectors
$\vec{\alpha}_i$ and $m$ constants $\Lambda_i$ which can have
positive or negative signs. The $\vec{\alpha}_i$ vectors form an
$m\times N$ matrix $\alpha_{iI}$, where the indices $i=1,\ldots,m$
 parametrise the exponential terms in the potential and the
indices $I=1,\ldots,N$  parametrise the different scalars. The
Lagrangian for the system then reads\footnote{We use the
convention for the metric with signature mostly plus.}:
\begin{equation} \label{lagrangian}
\mathcal{L}=\sqrt{-g}\,\Big(
R-\tfrac{1}{2}\,(\partial\vec{\phi})^2 - V\vec{(\phi)}\Big)\,.
\end{equation}
The equations of motion derived from the Lagrangian are
\begin{align} \label{EOM1}\nonumber
& \ddot{\phi}_I + 3H\dot{\phi}_I + \frac{\partial V}{\partial
 \phi_I}=0\,,\\
& H^2=\tfrac{1}{12}\,(\dot{\vec{\phi}}\cdot\dot{\vec{\phi}}) +
\tfrac{1}{6}\,V \,,\\\nonumber &
\dot{H}=-\tfrac{1}{4}\,(\dot{\vec{\phi}}\cdot\dot{\vec{\phi}})\,,
\end{align}
where the dot is differentiation w.r.t.~cosmic time. We refer to
the equations  as the scalar equations, the Friedmann equation and
the acceleration equation, respectively. The Hubble constant $H$
is defined as $H=\dot{a}/a$ where $a(\tau)$ is the scale factor
appearing in the flat FLRW metric:
\begin{equation}\label{mjetric}
ds^2=-d\tau^2+a(\tau)^2\,dx_3^2\,.
\end{equation}
There are $N+1$ degrees of freedom, namely, the scale factor and
the $N$ scalars (and accordingly only $N+1$ equations of motion
are independent. For example, the acceleration equation can be
obtained from the Friedmann equation and the scalar equations).
There exist 2 types of solutions:
\begin{itemize}
\item Critical points: These solutions correspond to stationary
solutions defined in terms of certain dimensionless variables,
which will be introduced in the next section. The critical points
can be obtained explicitly and they correspond to power-law
solutions ($a(\tau) \sim \tau^p$) or de Sitter
solutions\footnote{Anti-de Sitter solutions are not possible since
a flat FLRW metric doesn't support them} ($a(\tau)\sim e^\tau$).
The solutions can be either attractors, repellers or saddle
points. In the former two cases they correspond to the asymptotic
behaviour of more general solutions, whereas a saddle point just
corresponds to an intermediate regime. \item Interpolating
solutions: These are the non-stationary solutions and in general
they will interpolate between the critical points. Often they can
not be found explicitly, but a numerical analysis can reveal most
of their properties.
\end{itemize}

\section{The Critical Points}
Critical points (also known as fixed points or equilibrium points)
are solutions of differential equations in the context of
autonomous dynamical systems. An autonomous system is defined as a
system described by $n$ variables, say $\vec{z}$, whose dynamical
equations are of the form:
\begin{equation}\label{definition autonomous}
\frac{d\vec{z}}{dt}=\vec{f}(\vec{z})\, ,
\end{equation}
where $\vec{f}:\mathbb{R}^n \rightarrow \mathbb{R}^n$ is
interpreted as a vector field on $\mathbb{R}^n$. The differential
equations then imply that the vector field $\vec{f}$ is everywhere
tangent to the possible orbits. Critical points of an autonomous
system are defined as those points  $\vec{z}_0$ obeying
$\vec{f}(\vec{z}_0)=0$. These points are always exact constant
solutions since $d{\vec{z}}_0(t)/dt=0$. The nice property about
these systems is that the critical points are often the end points
(and initial points) of the orbits and therefore describe the
asymptotic behaviour. If the solutions interpolate between
critical points they can be divided into two classes:
\begin{itemize}
\item Heteroclinic orbit: This is an orbit connecting two
different critical points. \item Homoclinic orbit: This is an
orbit connecting a critical point to itself.
\end{itemize}
Most of the examples we found are of the first type and we will
focus on these. More on the theory of dynamical systems in
cosmology can be found in \cite{ellis:1997,Coley:1999uh}.

A nice property of multi-scalar cosmology with exponential
potentials is that they allow for a description in terms of
variables that make the system autonomous
\cite{Halliwell:1987ja,Copeland:1998et,Guo:2003eu,Coley:1999mj}.
With an arbitrary multi-exponential potential, the variables are
defined as follows:
\begin{equation}
x_{I}=\frac{\dot{\phi}_{I}}{\sqrt{12}\,H}\,, \,\,\,\,\,
y_{i}=\sqrt{\frac{\Lambda_{i}\,e^{-\vec{\alpha}_{i}\cdot\vec{\phi}}}{6H^2}}\,.
\label{x,y-coordinaten}
\end{equation}
In this notation, there are $N+m$ variables. Note that $y_i$ will
be imaginary when $\Lambda_i<0$, but this is not a problem since
only $y_i^2$ appears in the equations of motion. Rewriting the
equations of motion with these variables yields
\begin{align}
&  \frac{\dot{x}_I}{H}= - 3\,y^2\,x_{I} + \sqrt{3}\,\sum_{i=1}^m
\alpha_{iI}y^2_i\,,  \label{scalareq.2}\\  & x^2 + y^2 = 1\,,
\label{Friedmanneq2.}\\ & \frac{\dot{H}}{H^2}=-3\,x^2\,,
\label{acceleratingeq2.}
\end{align}
where we have used the shorthand notation $x^2=\sum^N_{I=1} x^2_I$
and $y^2=\sum^m_{i=1}y^2_i$. A nice consequence of the choice of
variables is that the Friedmann equation (\ref{Friedmanneq2.})
becomes the defining equation of an ($N+m-1$)-sphere (for
$\Lambda_i>0$, otherwise it will be a generalised hyperboloid).
Furthermore, from the acceleration equation, it is easily seen
that the condition for accelerate expansion is
\begin{align}
\ddot{a}>0\qquad \Leftrightarrow \qquad x^2<\frac{1}{3}\,.
\end{align}
The above condition allows us to visualise the region of
acceleration for the specific examples in section 4 and 5.

It turns out that we also need the derivatives of the
$y$-variables:
\begin{equation}\label{y-derivatives}
\frac{\dot{y}_{i}}{H}=\sqrt{3}\,(\sqrt{3}\,x^2-\vec{\alpha}_{i}\cdot
\vec{x})\,y_{i}\,.
\end{equation}
We can also use $\ln(a)$ as evolution parameter\footnote{One has
to be careful if the scale factor is not strictly monotonous.}
instead of cosmic time and it simplifies the equations since $H$
drops out in the scalar equations of motion and the equations for
$\dot{y}_i$, giving
\begin{align}\label{autonomous1}
\boxed{\  x_I'= - 3\,y^2\,x_{I} + \sqrt{3}\,\sum_{i=1}^m
\alpha_{iI}y^2_i\,,\quad
y_i'=\sqrt{3}\,(\sqrt{3}\,x^2-\vec{\alpha}_{i}\cdot
\vec{x})\,y_{i}\,,}
\end{align}
where the prime indicates differentiation w.r.t.~$\ln(a)$. The
above is clearly of the form (\ref{definition autonomous}), and
the critical points can therefore be calculated as $x_I'=y_i'=0$
(or equivalently as $\dot{x}_I=\dot{y}_i=0$). It is easy to prove
that the system will obey the Friedmann constraint
(\ref{Friedmanneq2.}) at all times as long as it does so
initially. So, if we impose (\ref{Friedmanneq2.}) on the initial
conditions then (\ref{autonomous1}) contains all information of
the subsequent evolution.

Integrating the acceleration equation (\ref{acceleratingeq2.}) for
a critical point yields power-law solutions  if $x^2 \neq 0$
\begin{equation}\label{power}
a(\tau) \sim \tau^p,\qquad  p=\frac{1}{3\,x^2}\,.
\end{equation}
If on the other hand $x^2=0$ we are in an extremum of the
potential with a de Sitter expansion
\begin{equation}
a(\tau) \sim \exp(\sqrt{\tfrac{1}{6}V(\phi_c)}\,\tau)\,.
\end{equation}

The equations \eqref{autonomous1} determining the critical points
are
\begin{align}
&(\sqrt{3}\,x^2-\vec{\alpha}_{i}\cdot\vec{x})\,y_i=0 \label{CP1}\,, \\
&-3\,y^2\,x_{I} + \sqrt{3}\,\sum_{i=1}^m
\alpha_{iI}y^2_i=0\,.\label{CP2}
\end{align}
There are two different kinds of critical points: \vskip3mm
\begin{tabular}{llll}
$\bullet$ &  The proper solutions: & & $\hspace{1cm}\slashed{\exists} i: y_i=0\,,$ \\
$\bullet$ &  Non-proper solutions: && $\hspace{1cm}\exists i: y_i=0\,.$   \\
\end{tabular}
\vskip3mm \noindent We can single out special non-proper
solutions, which always exist, namely the case where all $y$'s
vanish. From the Friedmann equation it follows that these
solutions have $x^2=1$ and for this reason we refer to them as
``the equator''. The solutions with some $y$'s vanishing have
infinite scalars and are therefore not proper solutions of the
equations of motion, but they are important as asymptotic
behaviour of interpolating solutions. From this classification we
see that there are a maximum of $2^m$ types of critical point
solutions \cite{Guo:2003eu}. Below we will give these solutions
for the most general exponential potential by analysing
(\ref{CP1}) and (\ref{CP2}).
\\

The rank $R$ of the matrix $\alpha_{iI}$, i.e.~the number of
independent $\vec{\alpha}_i$-vectors, plays a central role in this
discussion. In fact, the discussion of the general potential
naturally splits up into two cases: $R=m$ and $R<m$.

The rank $R$ gives the effective number of scalars appearing in
the potential, corresponding to the part of the scalar space that
is projected on the $\vec{\alpha}_i$-vectors. It is therefore
always possible to perform a field redefinition, such that only
$R$ scalars appear in the potential. The part of the scalar space
perpendicular to the $\vec{\alpha}_i$-vectors only appears in the
kinetic term of the Lagrangian and is ($N-R$)-dimensional. These
scalars therefore decouple from the rest. All systems with $N>R$
 have decoupled scalars and this is necessarily the case when
 $N>m$. Systems with $N
\leq m$ only have decoupled scalars if the vectors
$\vec{\alpha}_i$ are linearly dependent in such a way that $N>R$.

The field redefinition yielding $R$ scalars in the potential can
be performed by an $SO(N)$ rotation (which leaves the kinetic term
invariant) such that $\vec{\phi}$ changes into $\vec{\phi'}$ and
$\alpha'_{iR+1}=\alpha'_{iR+2}=\ldots=\alpha'_{iN}=0$ for all $i$.
We then notice from (\ref{CP2}) that for critical points all $x$'s
corresponding to decoupled scalars are zero, $x_{R+1}=x_{R+2}=
\ldots= x_N=0$. So in the rest of this section, the indices $I$
now run from 1 to $R$. In the case $R=m$, this makes $\alpha_{iI}$
a square matrix.

We have seen that the discussion of the system can be split up
into two cases, depending on the rank of $\alpha_{iI}$.
Alternatively, we can formulate this in terms of the following
matrix, which is quadratic in the $\alpha$'s
\begin{align}\label{amatrix}
A_{ij}=\vec{\alpha}_i \cdot \vec{\alpha}_j\,.
\end{align}
The separation of the general exponential potential into two
classes can then be characterised by the determinant of $A$:
\begin{align}
& R=m :\qquad {\rm det}(A) \neq 0\,, \label{R=m_class}\\\nonumber
& R<m :\qquad {\rm det}(A) = 0 \,.\label{R<m_class}
\end{align}
The first class corresponds to an invertible $A$-matrix and this
is exactly what is termed generalized assisted inflation
\cite{Copeland:1999cs}, whereas the second class, to our
knowledge, has not been treated in generality in the literature.

We will extend the existing results by also treating the case of
non-invertible $A$ in generality and by giving also the non proper
critical points of both classes. Special examples can be obtained
by performing compactifications over certain three-dimensional
unimodular group manifolds, corresponding to class A in the
Bianchi classification see e.g. \cite{Bergshoeff:2003vb}.

There is a subtlety about the description in terms of the
($x_I,y_i$)-variables, namely if $R<m$ then the $y$-variables are
not necessarily independent. We will comment on this in section
3.2.

\subsection{The $R=m$ Case}

A nice feature about this case is that $\dot{x}_I=0$ implies
$\dot{y}_i=0$. This can be seen in the following way: First we
differentiate (\ref{scalareq.2}) and use
$\dot{x}_I=d(y^2)/d\tau=0$. Multiplying with $\alpha_{jI}$ and
summing over $I$ we get $\sum_j(A_{ij})\,d(y^2_j)/d\tau=0$, and
since ${\rm det}(A) \neq 0$ we know that the only solution is
$d(y^2_i)/d\tau=0$.\\

We will now solve for the critical points:
\begin{itemize}\item \underline{Proper critical points}. From (\ref{CP1}) and (\ref{CP2}) we get:
\begin{equation}\label{square1}
\sum_i(A_{ij})\,y^2_i=3y^2\,x^2\,e_j\,,
\end{equation}
where $e_j$ is an $m$-dimensional vector with all components equal
to 1. Inverting this relation and using \eqref{CP2} yields the
values of $y_i$ and $x_I$ for the proper critical point
\begin{align}\label{bla1}
y^2_i=\frac{3p-1}{3p^2}\,\sum_{j=1}^m (A^{-1})_{ij}\,,\qquad
x_I=\frac{\sqrt{3}\,p}{3p-1}\,\sum_{i}^m\alpha_{iI}\,y^2_i\,,
\end{align}
where $p$ is the exponent given in \eqref{power}. The result for
$x_I$ can also be given in the rotated basis where $\alpha_{iI}$
is a square matrix
\begin{align}
x_I=\frac{1}{\sqrt{3}\,p}\,\sum_{i=1}^m(\alpha^{-1})_{iI}\,.
\end{align}
Note that by construction the $A_{ij}$-matrix \eqref{amatrix} is
$SO(N)$-invariant and accordingly all quantities containing only
this matrix can be calculated in any basis. We notice from our
formula above that there is a unique proper critical point.
However, it only exists when $y_i^2$, determined from
\eqref{bla1}, has the same sign as $\Lambda_i$, which serves as a
consistency check of definition \eqref{x,y-coordinaten}. Thus,
this critical point only exists for certain values of the
$\alpha$-vectors.

Using \eqref{power} we get the exponent for the power-law, which
reproduces the result found in
\cite{Copeland:1999cs,Green:1999vv}:
\begin{equation}\label{powerlaw1}
p=\sum_{i,j=1}^m(A^{-1})_{ij}\,.
\end{equation}
By integration we can go back to the $\phi_I,H$ variables where
the solution becomes:
\begin{align}
 H=\frac{p}{\tau}\,,\qquad
 \phi_I=\sqrt{12}\,p\ x_I\,\ln(\tau) + c_I\,,\qquad
 y^2_i=\frac{k_i}{\tau^2}\,, \label{Copelandansatz}
\end{align}
where $c_I$ and $k_i$ are integration constants. In fact, in
\cite{Copeland:1999cs,Chen:2003dc}, (\ref{Copelandansatz}) was
used as an Ansatz to find power-law solutions. \item
\underline{Non-proper critical points}. These correspond to some
$y$'s being equal to zero. Parametrising the subset of nonzero
$y$'s with the indices $a,b,c,\ldots$, the equations become:
\begin{align}
\sqrt{3}x^2-\vec{\alpha}_a\cdot\vec{x}=0\,,\qquad
\sum_{a}\alpha_{Ia}(y_a)^2-(1-x^2)\sqrt{3}x_{I}=0\,,
\end{align}
from which we deduce:
\begin{equation}\label{CP3}
\sum_b(A_{ab})y^2_b=3\,y^2\,x^2\,e_a\,.
\end{equation}
The $\vec{\alpha}_a$-vectors are of course also linearly
independent and accordingly the sub-matrix $A_{ab}$ has non-zero
determinant and is invertible.  Inverting relation (\ref{CP3}) and
using \eqref{CP2}, we find a unique solution
\begin{align}\label{bla2}
y^2_a=\frac{3p-1}{3p^2}\,\sum_{b} (A^{-1})_{ab}\,,\qquad
x_I=\frac{\sqrt{3}\,p}{3p-1}\,\sum_{a}^m\alpha_{aI}\,y^2_a.
\end{align}
The power-law is again given by (\ref{powerlaw1}) but now with the
inverse of the sub-matrix $A_{ab}$. Just as for the proper
solution, the above is only well-defined when $y_a^2$ has the same
sign as $\Lambda_a$. Note that all the above formulae for the
non-proper critical points are similar to those for the proper
ones. This is a consequence of the fact that vanishing $y$'s just
yield a truncated potential.
\end{itemize}

Note that, since the solution for the proper critical point is
unique and has power-law behaviour for the scale factor, there are
no de Sitter solutions. This can also be seen from
\eqref{square1}. Since $A$ has maximal rank, this matrix only has
the trivial nullspace, i.e $y_i=0$, which is not consistent with
the Friedmann equation, since $x=0$ for the de Sitter solutions.
We can conclude that potentials with linearly independent
$\vec{\alpha}_i$-vectors generically have power-law solutions and
no de Sitter solutions. This conclusion was also reached in
\cite{Ivashchuk:2003dw} where special cases were considered.

The special case where $\alpha_{iI}$ is diagonalisable by an
$SO(N)$-rotation is equivalent to the case where just one scalar
appears in each exponential, thus yielding the model which has
been called assisted inflation \cite{Liddle:1998jc}.

\subsection{The $R<m$ Case}

Since ${\rm det}(A)=0$ we will have to use another approach to
determine the critical points. And also the $R<m$ case will be
more difficult to treat in full generality because the $y$'s are
not necessarily independent.

The number of independent $y$'s is always smaller than or equal to
$R+1$, as we will now illustrate. After possible field
redefinitions, the $y$-coordinates are given in terms of $R+1$
fields, namely the scalars and the Hubble parameter. Among the $m$
coordinates we therefore at most have $R+1$ independent, e.g.
$y_1,\ldots,y_{{\scriptscriptstyle{R+1}}}$, and this leaves us
with $m-R-1$ relations for the rest of the $y$'s. From the
definitions of the $y$'s, we can express $\phi_I$ and $H$ in terms
of the first $R+1$ $y$'s
\begin{align}
e^{\phi_I}=\prod_{i=1}^{R}\Big(\frac{y_i^2\,\Lambda_{i+1}}{y_{i+1}^2\,\Lambda_i}\Big)^{(\beta^{-1})_{Ii}}\,,\qquad
H=\frac{\Lambda_{{\scriptscriptstyle
R+1}}}{6}\,e^{-\vec{\alpha}_{{\scriptscriptstyle{R+1}}}\cdot\vec{\phi}}\,y^{-2}_{{\scriptscriptstyle{R+1}}}\,,
\end{align}
where the following square matrix has been defined
\begin{align}
\beta_{iJ}=\alpha_{i+1,J}-\alpha_{iJ}\,,\quad
i,J\in\{1,\ldots,R\}\,.
\end{align}
We can then express the remaining $y$'s in terms of the first
$R+1$ as follows
\begin{align}
y_i^2=y^{2}_{{\scriptscriptstyle{R+1}}}\,\frac{\Lambda_i}{\Lambda_{{\scriptscriptstyle{R+1}}}}\frac{\prod_{j,K=1}^R
\Big(\frac{y_j^2\,\Lambda_{j+1}}{y_{j+1}^2\,\Lambda_j}\Big)^{\alpha_{iK}(\beta^{-1})_{Kj}}}
{\prod_{l,M=1}^R\Big(\frac{y_l^2\,\Lambda_{l+1}}{y_{l+1}^2\,\Lambda_l}\Big)^{\alpha_{R+1,M}(\beta^{-1})_{Ml}}}\,,\qquad
i=R+2,\ldots,m\,.
\end{align}
Thus, the maximal number of independent $y$'s is $R+1$. It is
possible to prove that the above relations for the $y_i$'s will be
obeyed for the dynamical system \eqref{autonomous1} at all times
if they do so initially. So again we can use \eqref{autonomous1}
as equations which govern the whole system as long as we pick our
initial conditions consistent. With this in mind we will look for
critical points.

Until now we denoted the row vectors of the $\alpha$-matrix with
$\vec{\alpha}_i$ and $A_{ij}$ is the matrix with as entries the
inner products of these row vectors: $A_{ij}=\vec{\alpha}_i\cdot
\vec{\alpha}_j$. In this section we will also need the column
vectors which we will denote by $\vec{\alpha}_I$ and we then
define the following matrix
\begin{align}
B_{IJ}=\vec{\alpha}_I\cdot \vec{\alpha}_J\,.
\end{align}
 The $R$ column vectors $\vec{\alpha}_I$ are
all linearly independent because the rank of $\alpha$ equals $R$
and as a consequence $B$ is invertible (remember that $I$ now runs
from 1 to $R$). It is this property that we will use to find the
solutions.
\begin{itemize}\item \underline{Proper power-law critical points}.
Looking for the solution(s), with $y_i \neq 0$, we find
from (\ref{CP1})
\begin{equation}\label{square2}
\sum^R_{I=1}B_{IJ}x_I=\sqrt{3}x^2F_J\,,
\end{equation}
where $F_J=\sum^m_{i=1}\alpha_{iJ}$. Thus we can solve for $x_I$:
\begin{equation}\label{xsolution2}
x_I=\frac{1}{\sqrt{3}\,p}\,\sum^R_{J=1}(B^{-1})_{IJ}\,F_J\,,
\end{equation}
and hence we find the extension of the power-law formula to the
case where $R<m$:
\begin{equation}\label{powerlaw2}
p=\vert B^{-1}\cdot \vec{F}\vert^2\,.
\end{equation}
One can prove that this formula reduces to (\ref{powerlaw1}) if
$R=m$. Since the rank of $\alpha_{iI}$ is $R$, it is enough to use
$R$ independent equations among the $m$ equations of \eqref{CP1}
to obtain $x_I$. This result of course has to be consistent with
the remaining $m-R$ equations, and this puts strong restrictions
on the allowed dilaton couplings as we will now show. Let
$\{\vec{\alpha}_a\}_{a=1}^R$ be linearly independent. It is
possible to solve \eqref{CP1} simultaneously for these vectors.
The rest of the vectors can be written as linear combinations and
are only guaranteed to solve \eqref{CP1} if the linear
combinations are convex \footnote{Of course there are many ways to
number the vectors; it is enough to find one which obeys these
relations.}
\begin{align}\label{convex}
\vec{\alpha}_i=\sum_{a=1}^R c_{ia} \vec{\alpha}_a\,,\qquad
\sum_{a=1}^R c_{ia}=1\,,\qquad i=R+1,\ldots,m\,.
\end{align}
We will give a specific example, which has an M-theory origin,
where this is realised. A special case is $R=1$, where after field
redefinitions only one scalar appears in the potential. In this
case, the above solution will never exist, since \eqref{CP1}
becomes $m$ equations with one variable (or equivalently, the
requirement of convexity here would imply $m=1$ which is not the
case under consideration).

An important difference from the previous case is the question of
the uniqueness of the solution. We can not obtain the $y$-values
with this procedure, and in particular we cannot determine whether
they are unique. In fact, it is easy to give an example where they
are not: When at least one $\Lambda_i<0$ we have the following
possibility, since $A$ has non-trivial kernel
\begin{align}\label{yline}\nonumber
y^2&=0\,,\quad y_i^2\in {\rm Ker}(A)\,,\\
x^2&=1\,,\quad \vec{\alpha}_i\cdot\vec{x}=\sqrt{3}\,,\quad {\rm
for}\quad y_i\neq 0\,.
\end{align}
In particular, this includes a proper critical point of the form
\eqref{xsolution2} when all $y_i\neq 0$ and where furthermore
\begin{align}
\vert B^{-1}\cdot \vec{F}\vert^2=\frac{1}{3}\,.
\end{align}

\item \underline{De Sitter solutions}. It was seen in the previous
subsection, that de Sitter solutions do not exist for $R=m$,
because the matrix $A$ then has a trivial kernel. In the present
case, since $A$ has a non-trivial kernel, and therefore a de
Sitter solution is possible
\begin{align}\label{desitter}
x=0\,,\quad y^2=1\,,\quad y^2_i\in {\rm Ker}(A)\,.
\end{align}
Again, this solution is only well-defined when $y_i^2$ has the
same sign as $\Lambda_i$. We can conclude that potentials with
$R<m$ show the opposite behaviour of $R=m$ potentials. Here
(proper) power-law solutions are rare (only possible for certain
couplings \eqref{convex}) whereas de Sitter solutions are quite
generic. Again, a similar observation was made in
\cite{Ivashchuk:2003dw}, but for specific couplings (which did not
allow power-law behaviour).

\item \underline{Non-proper critical points}. Looking for these
solutions, we again put a subset of the $y$'s to zero. This
corresponds to some terms in the potential being absent and we can
therefore analyse the new system as before but with a
``truncated'' potential. A subtlety is that whenever $R<m$ the
$y$'s are dependent on each other, and therefore only certain
subsets of the $y$'s can be zero at the same time.

\end{itemize}

The findings of this section are summarised in the table below.
The asterisk in the lower left corner symbolises the fact that we
have a truncated system which can belong to either of the two
cases ($R<m$ or $R=m$).

\begin{table}[ht]
\begin{center}
\hspace{-1cm}
\begin{tabular}{||c|c|c||}
\hline \rule[-1mm]{0mm}{6mm}
   & $R<m\,,\quad {\rm det}(A)=0$ & $R=m\,,\quad {\rm det}(A)\neq 0$ \\
\hline \hline \rule[-1mm]{0mm}{6mm}
 Proper & Power-law (convex combinations) & Power-law\\ \cline{2-3}
& de Sitter & No de Sitter \\
\hline\rule[-1mm]{0mm}{6mm} Non-proper & $\ast$ & Power-law
\\ \cline{3-3} &  & No de Sitter \\ \hline
\end{tabular}
\caption{\it The critical points for multi-exponential
potentials.}
\end{center}
\end{table}

As mentioned, the critical points give rise to the asymptotic
behaviour of the general solutions. By performing a stability
analysis it is possible to determine the nature of the critical
points, i.e.~whether they are attractors, repellers or saddle
points. This can be done by linearising the system around the
critical points, $\vec{x}'={\bf M}\cdot\vec{x}$, and determining
the eigenvalues of the matrix ${\bf M}$. If the real part of all
eigenvalues is negative, the critical point is an attractor, if
the real part of all eigenvalues is positive, the critical point
is a repeller and in the mixed case it is a saddle point. It is
easy to perform the stability analysis in the simple cases
considered in the following sections and the result is confirmed
by the interpolating solutions, which are calculated numerically.

\section{Double and Triple Exponential Potentials}

In this section we will consider some specific examples of
 double and triple exponential potentials with one or two scalars, i.e.~$m=2, 3$ and $N=1, 2$. These
examples serve as an illustration of the formal framework in the
previous section.

As mentioned, the critical points give the asymptotic behaviour of
more general solutions. In some cases it has been possible to
obtain these solutions exactly. For single exponential potentials,
this has been done for arbitrary dilaton couplings and the result
can be pictured as straight lines in the space defined by the
$x$'s \cite{Bergshoeff:2003vb}. For double exponential potentials,
exact solutions have been obtained for special values of the
dilaton couplings, corresponding to the reduction of S-brane
solutions to 4D, see
e.g.~\cite{Ohta:2003pu,Roy:2003nd,Wohlfarth:2003kw} and references
therein. Ideally, we would like to obtain exact results for the
general case. However, this is a highly non-trivial task, and we
therefore turn to numerical methods, which can still show the
qualitative behaviour of the solutions. To this end, it is
convenient to use $\ln(a)$ as a time parameter. For an eternally
expanding universe where $a$ increases from 0 to $\infty$, our
time coordinate ranges from $-\infty$ to $\infty$.

In general, an S-brane can be obtained as a time-dependent
solution to the following system containing gravity, an
antisymmetric tensor and possibly a dilaton
\begin{align}
S=\int
d^{4+d}x\,\sqrt{-\hat{g}}\,\Big(\hat{R}-\tfrac{1}{2}\,(\partial\hat{\phi})^2-\tfrac{1}{2
d!}\,e^{-b\hat{\phi}}\,\hat{F}^2_d\Big)\,,
\end{align}
where the hats indicate that the fields live in $4+d$ dimensions
and where the dilaton coupling for maximal supergravities is given
by
\begin{align}
b=\sqrt{\frac{14-2d}{d+2}}\,.
\end{align}
Reducing over a $d$-dimensional maximally symmetric space with
curvature $k$ and flux $f$ yields the following potential
\cite{Bremer:1998zp}
\begin{align}\label{sbrane-pot}
V(\phi,\varphi)=f^2\,e^{-b\,\phi-3\sqrt{\frac{d}{d+2}}\,\varphi}-k\,e^{-\sqrt{\frac{d+2}{d}}\,\varphi}\,,
\end{align}
where $\varphi$ is the Kaluza-Klein scalar. S2-brane solutions
have been found in six to eleven dimensions, corresponding to
$d=2,\ldots,7$. In five dimensions, an S2-brane has a 1-form field
strength. The corresponding four-dimensional cosmological solution
with single exponential potential was found in
\cite{Bergshoeff:2003vb}. As explained in that paper, a general
twisted reduction leads to triple exponential potentials, which
could have corresponding exotic S-brane solutions in five
dimensions.

\subsection{Double Exponential Potentials, one Scalar}

The simplest case is $m=2$ and $N=1$. The corresponding potential
is then described in terms of two dilaton couplings $\alpha_1$ and
$\alpha_2$. We can always choose e.g.~$\alpha_1$ to be positive
and in this example we will start by considering positive
$\Lambda_i$. Since $R=1$, we have 2 independent $y$'s. The
Friedmann equation defines a 2-sphere, but the allowed solutions
can only lie on the part corresponding to non-negative $y$'s.
Using the machinery from the previous section, we find the
following critical points
\begin{alignat}{2}\label{critm2n1}
\nonumber (i)\qquad y_1&=y_2=0\,,\quad x^2=1\,, & &\\\nonumber
(ii)\qquad y_1&=\sqrt{1-\frac{\alpha_1^2}{3}},\quad y_2=0\,,\quad
x=\frac{\alpha_1}{\sqrt{3}}\,,  & {\rm for}&\quad \alpha_1^2<3 \\
(iii)\qquad y_1&=0\,,\quad
y_2=\sqrt{1-\frac{\alpha_2^2}{3}}\,,\quad
x=\frac{\alpha_2}{\sqrt{3}}\,, & {\rm for}&\quad \alpha_2^2<3
\\\nonumber
(iv)\qquad y_1&=(1-\frac{\alpha_1}{\alpha_2})^{-1/2}\,,  \quad
y_2=(1-\frac{\alpha_2}{\alpha_1})^{-1/2}\,,\quad x=0 \,, &
\quad{\rm for}&\quad \alpha_2<0\,.
\end{alignat}
The first corresponds to the ``equatorial'' points $x=\pm 1$. In
an $(y_1, y_2, x)$-plot these become the North and South Pole. The
proper solution only exists for $\alpha_2<0$ and then corresponds
to a de Sitter solution. The stability of the different points is
best illustrated by considering the different possible
cases\footnote{If we do not explicitly classify the stability of a
critical point, it will be a saddle point.}.

\begin{itemize}
\item $\alpha_1\,, \alpha_2>\sqrt{3}$: Only the North and South
Pole are critical Points; the former is attracting and the latter
is repelling and any interpolating solution will be a curve in
between, and these can be found numerically. An example is
illustrated in the left part of figure \ref{m2n1x}.

\item $\alpha_1<\alpha_2<\sqrt{3}$: The critical points (i)-(iii)
exist. The poles are repellers and (iii) is attracting.

\item $\alpha_1<\sqrt{3}\,,\alpha_2<-\sqrt{3} $: Apart from the
poles, we have the two critical points, corresponding to (iii) and
(iv) in \eqref{critm2n1}. The North Pole is repelling and the de
Sitter solution is attracting; this is shown on the right in
figure \ref{m2n1x}.

\begin{figure}[h]
\centerline{\epsfig{file=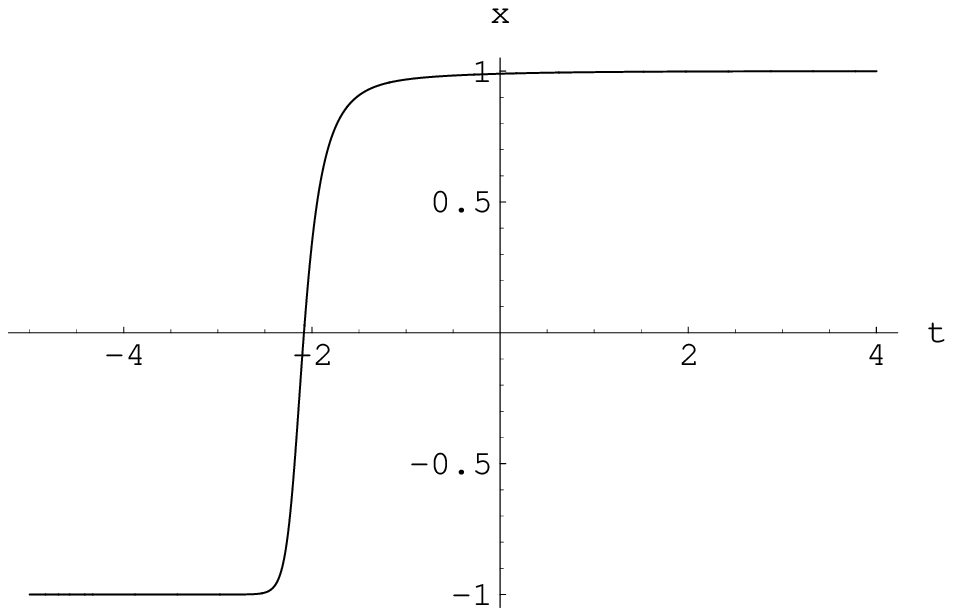,width=.4\textwidth}\hspace{1cm}
\epsfig{file=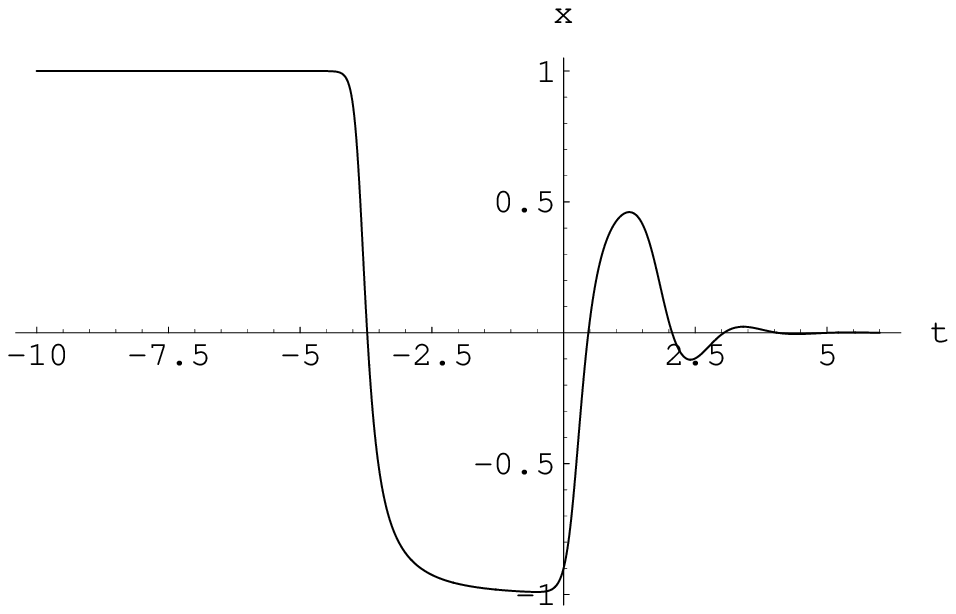,width=.4\textwidth}}
 \caption{\it {\small The plot to the left shows $x(t)$ in the case ($\alpha_1,\alpha_2)=(3,2)$,
 where the solution interpolates
 between the North and South Pole.
 The plot to the right is for the case $(\alpha_1,\alpha_2)=(1,-2)$, yielding
 a solution interpolating between the
North Pole and a de Sitter solution.}}
 \label{m2n1x}
\end{figure}

\item $\alpha_1>\sqrt{3}\,,-\sqrt{3}<\alpha_2<0$: This is similar
to the previous case, except that critical point (iii) is
interchanged with (ii), and the early asymptotics will be the
South Pole.

\item $\alpha_1>\sqrt{3}\,,0<\alpha_2<\sqrt{3}$: In addition to
the North and South Pole, there is the non-proper critical point
(ii), which is an attractor. The South Pole is repelling. An
interpolating solution is shown in the left part of figure
\ref{m2n1yyx}.

\item $\alpha_1>\sqrt{3}\,,\alpha_2<-\sqrt{3}$: The critical
points are the poles and the de Sitter solution, and the latter is
an attractor. It turns out that the poles are saddle points, and
hence they do not give rise to the early asymptotics of the
solution. Instead this will be an infinite cycle, moving closer
and closer to the boundary of the space (given by $y_1=0$ or
$y_2=0$) as time goes to minus infinity. This is illustrated to
the right in figure \ref{m2n1yyx}.

\begin{figure}[h]
\centerline{\epsfig{file=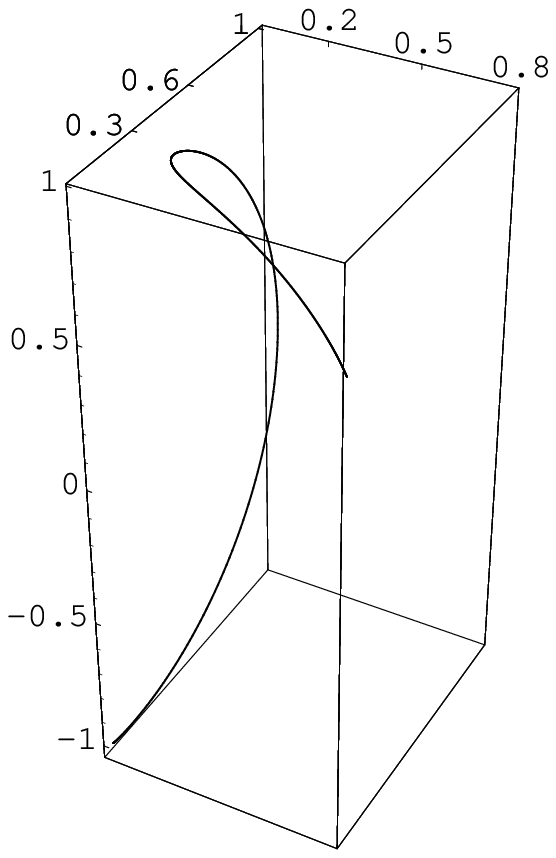,width=.2\textwidth}\hspace{2cm}
\epsfig{file=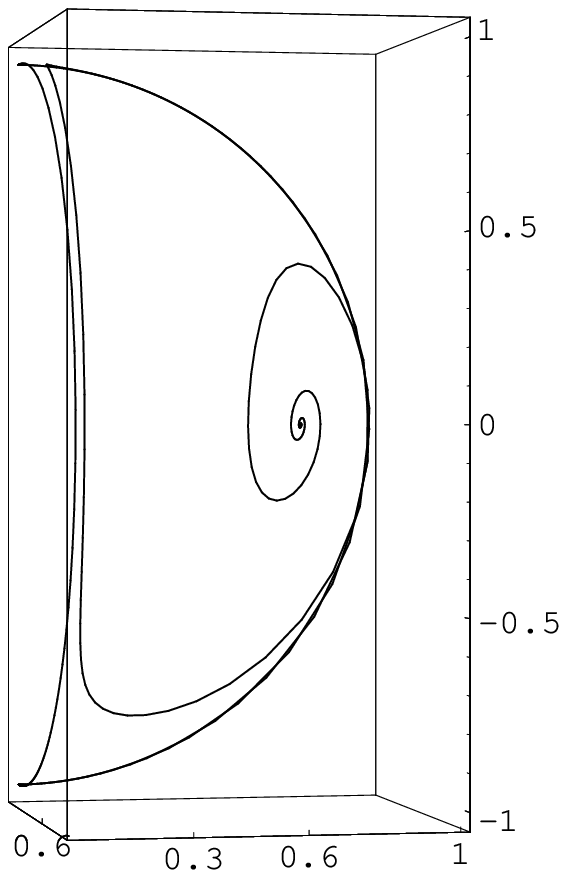,width=.2\textwidth}}
 \caption{\it {\small The figure shows $(y_1,y_2,x)$-plots for two cases. The left part, with
 $(\alpha_1,\alpha_2)=(2,1)$, shows a
  solution interpolating between the South Pole and the critical point (ii).
 The right plot, with $(\alpha_1,\alpha_2)=(3,-2)$ shows a solution spiralling towards the de Sitter point.}}
\label{m2n1yyx}
\end{figure}

\item $\alpha_1<\sqrt{3}\,,-\sqrt{3}<\alpha_2<0$: The late-time
asymptotics are similar to the previous case. The early-time
asymptotics are different due to the fact that all the critical
points (i)-(iv) are realised. Both of the poles will be repelling,
and depending on initial conditions either of these can give rise
to the early-time asymptotics.
\end{itemize}

For all the cases above, the solutions enter a phase of
acceleration when $x^2<1/3$. The cases with
$\vert\alpha_1\vert\,,\vert\alpha_2\vert>1$ give rise to one
period of transient acceleration, and otherwise the solution will
end up in a phase of eternal acceleration, which as mentioned is
an asymptotic de Sitter phase when $\alpha_2<0$. In the case
$\alpha_1>\sqrt{3}\,,\alpha_2<-\sqrt{3}$ the phase of late-time
acceleration is preceded by an infinite cycle, alternating between
acceleration and deceleration.

The case with $\Lambda_2<0$ can be analysed in a similar way, but
the interpolating solutions will now be given by curves on a
hyperboloid. The critical point $(iii)$ will now only exist for
$\alpha_2^2>3$, since this yields $y_2^2<0$. By the same token,
the de Sitter critical point $(iv)$ only exist for
$\alpha_2>\alpha_1>0$.

The S-brane case, corresponding to $\hat{\phi}=0$, gives the
following dilaton couplings
\begin{align}
\alpha_1=3\,\sqrt{\frac{d}{d+2}}\,,\quad
\alpha_2=\sqrt{\frac{d+2}{d}}\,.
\end{align}
This system, which can be obtained from eleven dimensions where it
will give rise to SM2-brane solutions, was analysed in
\cite{Jarv:2004uk}, where curvature of the external space is also
included. It is seen that only the critical points $(i)$ and
$(iii)$ exist for $\Lambda_2>0$ and $(i)$ and $(ii)$ exist for
$\Lambda_2<0$, with the latter being attracting. In the latter
case, we also have a de Sitter critical point which, however, is
not an attractor.

\subsection{Double Exponential Potential, two Scalars}

There is another case with double exponential potentials which has
two scalars, i.e.~$m=2$ and $N=2$. Considering the two
$\alpha$-vectors to be independent, we get $R=2$. The critical
points can be obtained as a special case of the general analysis
from the previous section and consist of the equator, $x^2=1$,
$y_i=0$, the proper critical point, $y_i\neq 0$ and two non-proper
critical points with $y_i=0\,,y_j\neq 0$, $i \neq j$.

\begin{figure}[h]
\centerline{\epsfig{file=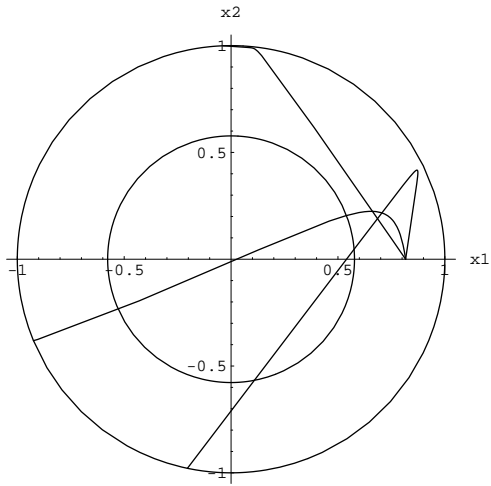,width=.4\textwidth}}\hspace{2cm}
 \caption{\it {\small Three interpolating solutions, corresponding to S2-branes reduced to four dimensions,
 projected on the $(x_1,x_2)$-plane. The inner circle is the boundary of the accelerating region.}}
\label{double}
\end{figure}

Specialising to the reduction of S-branes, we get the dilaton
couplings
\begin{align}
\vec{\alpha}_1=(3\,\sqrt{\frac{d}{d+2}},\sqrt{\frac{14-2d}{d+2}})\,,\quad
\vec{\alpha}_2=(\sqrt{\frac{d+2}{d}},0)\,.
\end{align}
For these $\alpha$-couplings we get ${\rm det}(A)=14/d-2$, and
therefore the matrix $A$ is invertible for $d<7$. However, using
\eqref{bla1}, $y_1^2$ is seen to be negative and since
$\Lambda_1=f^2>0$, the proper critical point does not exist. Apart
from the equator, there is another critical point, which has
$y_1=0$ and $y_2\neq 0$ and corresponds to a power-law behaviour
with exponent $p=d/(d+2)$. This critical point will be an
attractor and the equator will be a repeller. Thus, an S2-brane
reduced to four dimensions corresponds to a solution interpolating
between the equator and the attracting critical point. This is
similar to the behaviour of the solution found in
\cite{Townsend:2003fx}, which is the fluxless limit of a reduced
S2-brane \cite{Ohta:2003pu}. Indeed, the attracting power-law
solution is the same with or without flux. Examples of
interpolating solutions, projected on the $(x_1,x_2)$-plane,  are
shown in figure \ref{double} in the case of $d=2$. It is seen that
they indeed interpolate between the equator and the attracting
critical point, which according to \eqref{bla2} has the
coordinates $(x_1,x_2)=(\sqrt{2/3},0)$. For $d=7$ there is a
possibility of a de Sitter solution since ${\rm det}(A)=0$, see
\eqref{desitter}, but again it does not exist because $y_1^2$ is
negative.

\subsection{Triple Exponential Potential, one Scalar}

This example is the simplest case where the $y$-variables are not
all independent and this subsection serves as an illustration. The
potential is described in terms of three dilaton couplings
$\alpha_1$, $\alpha_2$ and $\alpha_3$. For simplicity, we take
$\Lambda_i>0$; the case with negative $\Lambda_i$ can be analysed
in a similar way. We can choose $\alpha_3>0$. Since $R=1$, we have
two independent $y$'s, leaving one relation, which reads
\begin{align}\label{yrel}
(\Lambda_2)^{\alpha_1-\alpha_3}\,(y^2_2)^{\alpha_3-\alpha_1}=(\Lambda_1)^{\alpha_2-\alpha_3}\,
(\Lambda_3)^{\alpha_1-\alpha_2}\,(y^2_1)^{\alpha_3-\alpha_2}\,(y^2_3)^{\alpha_2-\alpha_1}\,.
\end{align}
The analysis of critical points is analogous to the previous case,
except for the extra feature of the relation above.  There are
three kinds of critical points (for the moment we forget about the
$y$-dependence)
\begin{align}\label{critm3n1}\nonumber
(i)\qquad y_i&=0\,,\quad x^2=1\,,\\
(ii)\qquad y_i&=y_j=0\,,\ y_k=\sqrt{1-\frac{\alpha_k^2}{3}}\,,\
x=\frac{\alpha_k}{\sqrt{3}}\,,\quad i, j, k \quad{\rm
different}\\\nonumber (iii)\qquad x&=0 \,.
\end{align}
A necessary condition for its existence is $\alpha_k^2<3$.
However, this is not sufficient, since \eqref{yrel} only allows
certain $y$'s to be non-zero while the others are zero. For
instance, with $\alpha_3>\alpha_2>\alpha_1$, having $y_1=0$ or
$y_3=0$ implies $y_2=0$.

\begin{figure}[h]
\centerline{\epsfig{file=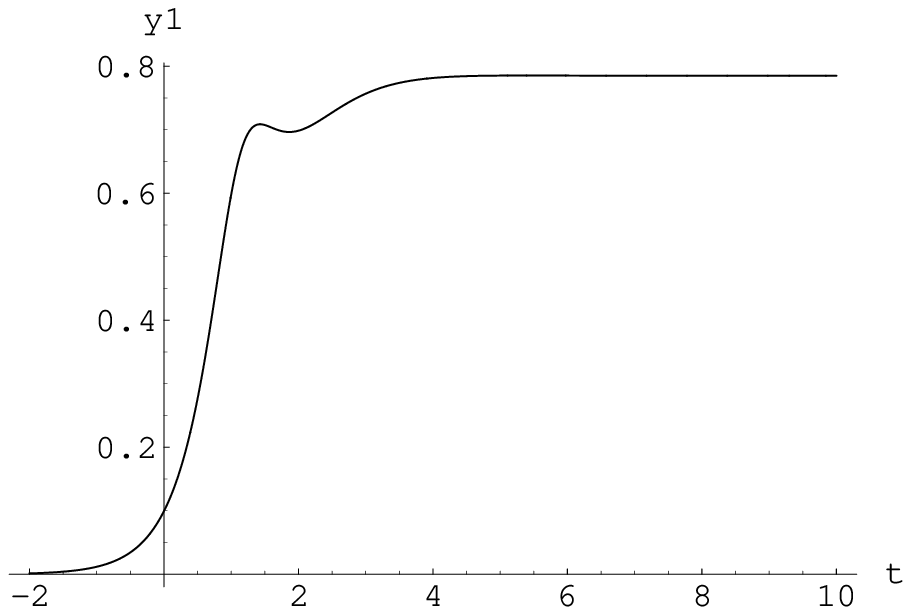,width=.3\textwidth}\hspace{.5cm}\epsfig{file=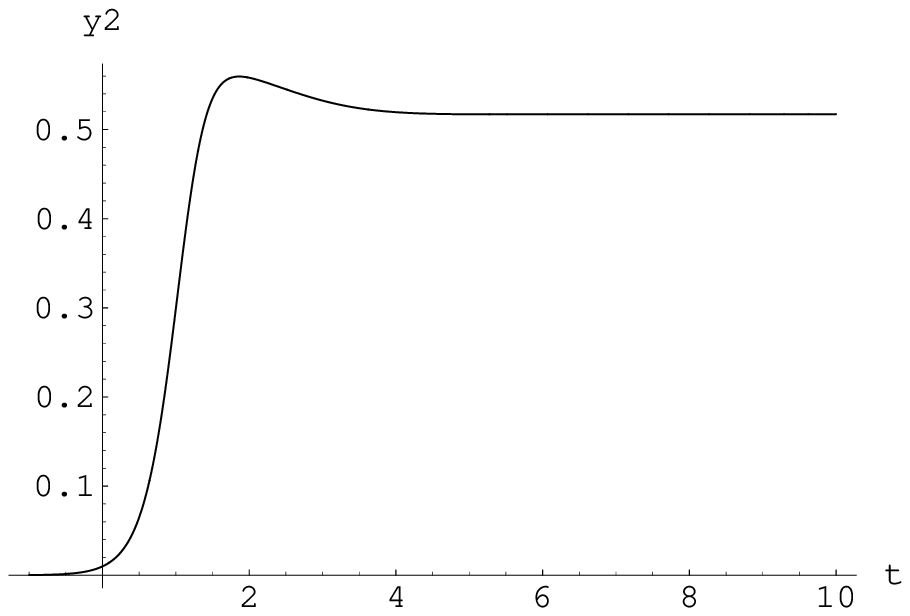,width=.3\textwidth}
\epsfig{file=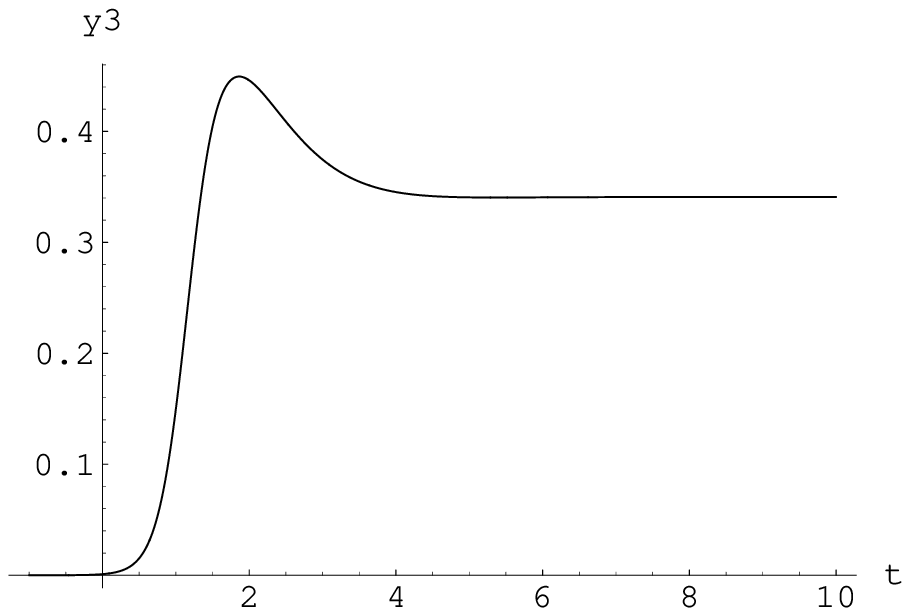,width=.3\textwidth}}
 \caption{\it {\small The plots show $y_1(t)$, $y_2(t)$ and $y_3(t)$, respectively. As $t$ increases, they
 tend towards the de Sitter critical point.}}
 \label{m3n1y}
\end{figure}

The third type of critical point is a de Sitter solution given by
the following equations
\begin{align}
\alpha_1\,y_1^2+\alpha_2\,y_2^2+\alpha_3\,y_3^2=0\,,\qquad
y_1^2+y_2^2+y_3^2=1\,,
\end{align}
which can be rewritten as
\begin{align}
\frac{\alpha_3-\alpha_1}{\alpha_3}\
y_1^2+\frac{\alpha_3-\alpha_2}{\alpha_3}\ y_2^2=1\,,
\end{align}
which defines an ellipse for $\alpha_3>\alpha_1\,,\alpha_2$. When
substituting $y_3$, \eqref{yrel} also gives a curve in the
$(y_1,y_2)$-plane, and the critical point is given as the
intersection between these two curves.\footnote{However, it is
only possible to give algebraic expressions of the solution for
special values of the dilaton couplings.} As an example, for
$(\alpha_1,\alpha_2,\alpha_3)=(-1/2,1/2,3/2)$ and $\Lambda_i=1$,
the de Sitter critical point becomes $(y_1=0.78, y_2=0.52,
y_3=0.34)$. Figure \ref{m3n1y} shows the time-development of an
interpolating solution for this case. It is seen that the
late-time behaviour indeed corresponds to the de Sitter critical
point above.

\section{Multi-exponential Potentials from Group Manifolds}

In this section we will consider specific cases, which can be
obtained by reducing pure gravity in seven dimensions over a
three-dimensional group manifold. Since pure gravity in 7D can be
embedded in 11D, the solutions have an M-theory origin. We will
focus on the triple-exponential case.

Double exponential potentials can be obtained for certain
truncations of reductions over type VIII and IX group manifolds
\cite{Bergshoeff:2003vb}. This is equivalent to a trivial
reduction over a circle followed by a reduction over a maximally
symmetric 2D space with flux. The resulting potential is given by
\eqref{sbrane-pot} with $d=2$, and interpolating solutions
correspond to reductions of S2-branes from six dimensions.

A triple exponential potential can be obtained from type VI$_0$
and VII$_0$ group manifolds and the result is
\cite{Bergshoeff:2003vb}
\begin{align}
V=\tfrac{1}{8}\,e^{-\sqrt{3}\varphi}\,(e^\phi\pm e^{-\phi})^2\,,
\end{align}
where the plus sign occurs for type VI$_0$ and the minus sign for
type VII$_0$. We therefore have an example with $m=3$ and $N=2$,
and the three dilaton couplings are
\begin{align}
\vec{\alpha}_1=(\sqrt{3},2)\,,\qquad\vec{\alpha}_2=(\sqrt{3},-2)\,,\qquad
\vec{\alpha}_3=(\sqrt{3},0)\,.
\end{align}
Note that only two of these are independent, and this case
therefore falls into the class with $R<m$, and more interestingly,
we find the convex combination
$\frac{1}{2}\vec{\alpha}_1+\frac{1}{2}\vec{\alpha}_2=\vec{\alpha}_3$,
so there is a possibility of a proper critical point with
power-law behaviour. The fact that we have linearly dependent
$\vec{\alpha}_i$-vectors ($R<m$) is actually the case for most
class A Bianchi types. For the present example, there will be two
independent $y$-variables plus the relation $y_3=\pm 2\,y_1 y_2$,
but only $y_1$ and $y_2$ are needed.

\begin{figure}[ht!]
\centerline{\epsfig{file=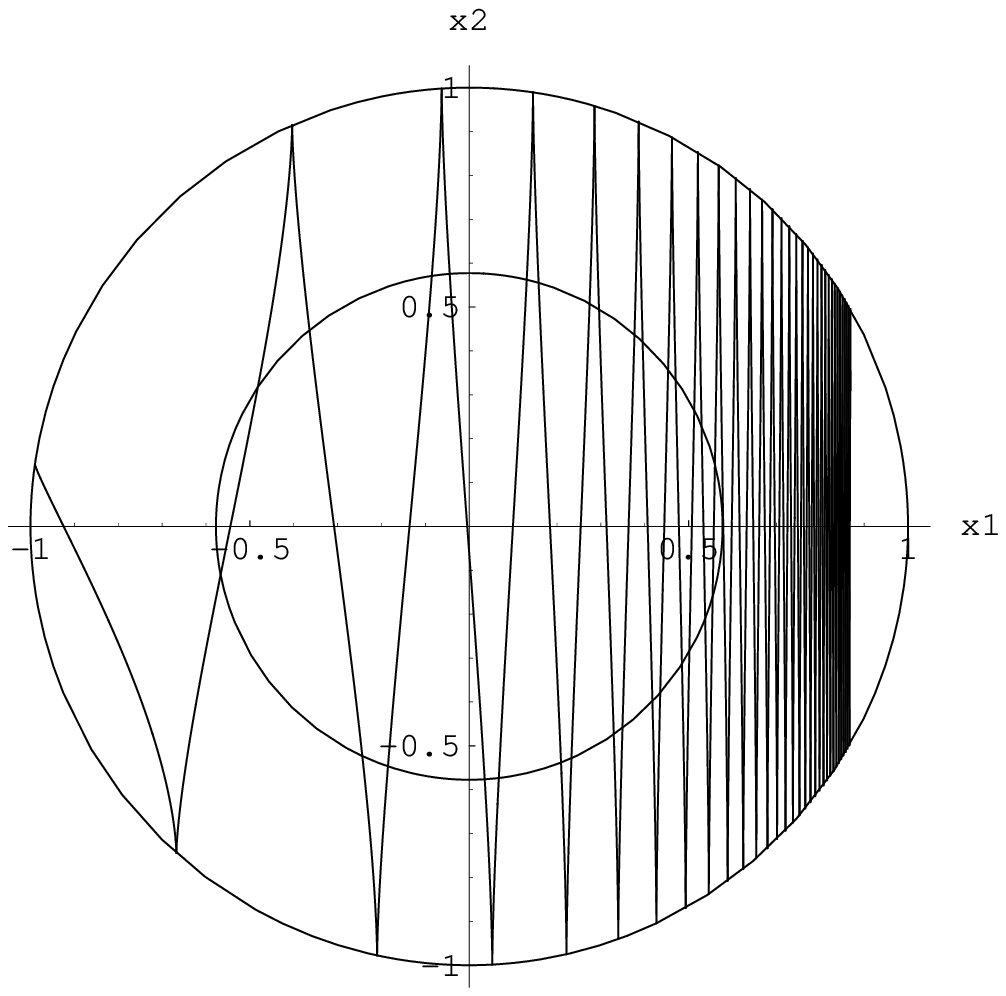,width=.23\textwidth}\hspace{1cm}
\epsfig{file=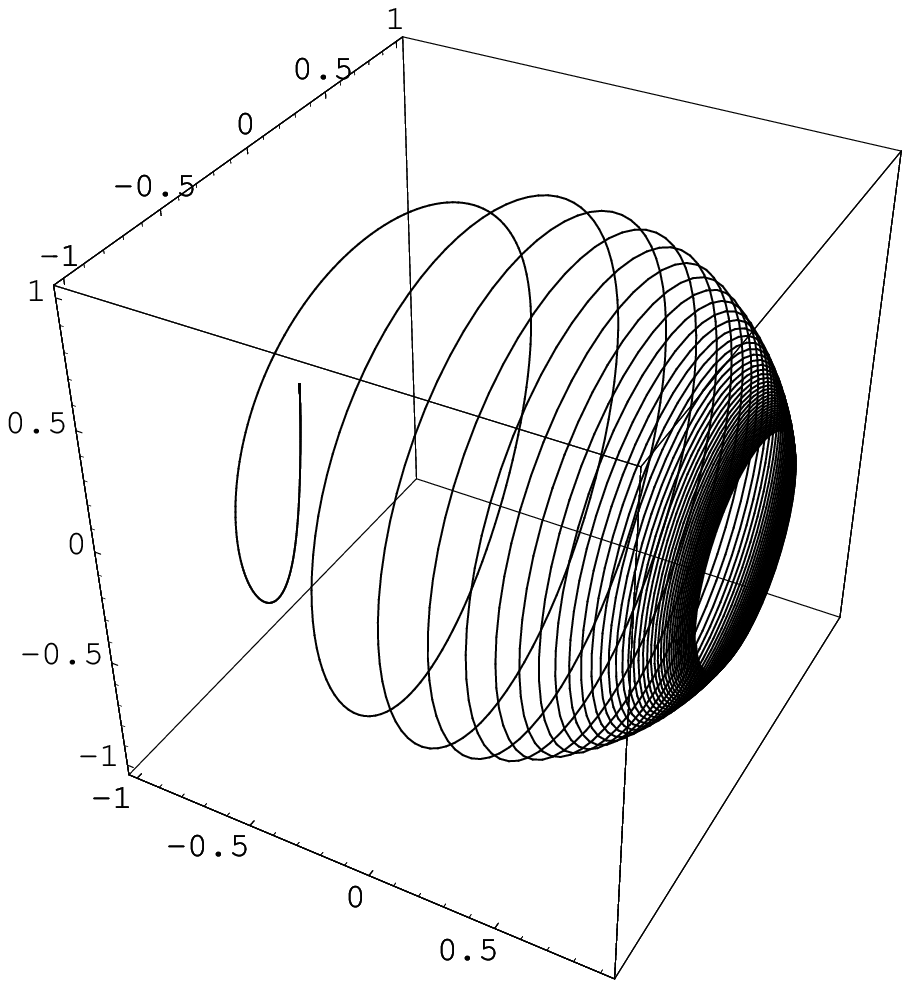,width=.23\textwidth}}
\vspace{0.5cm}
\centerline{\epsfig{file=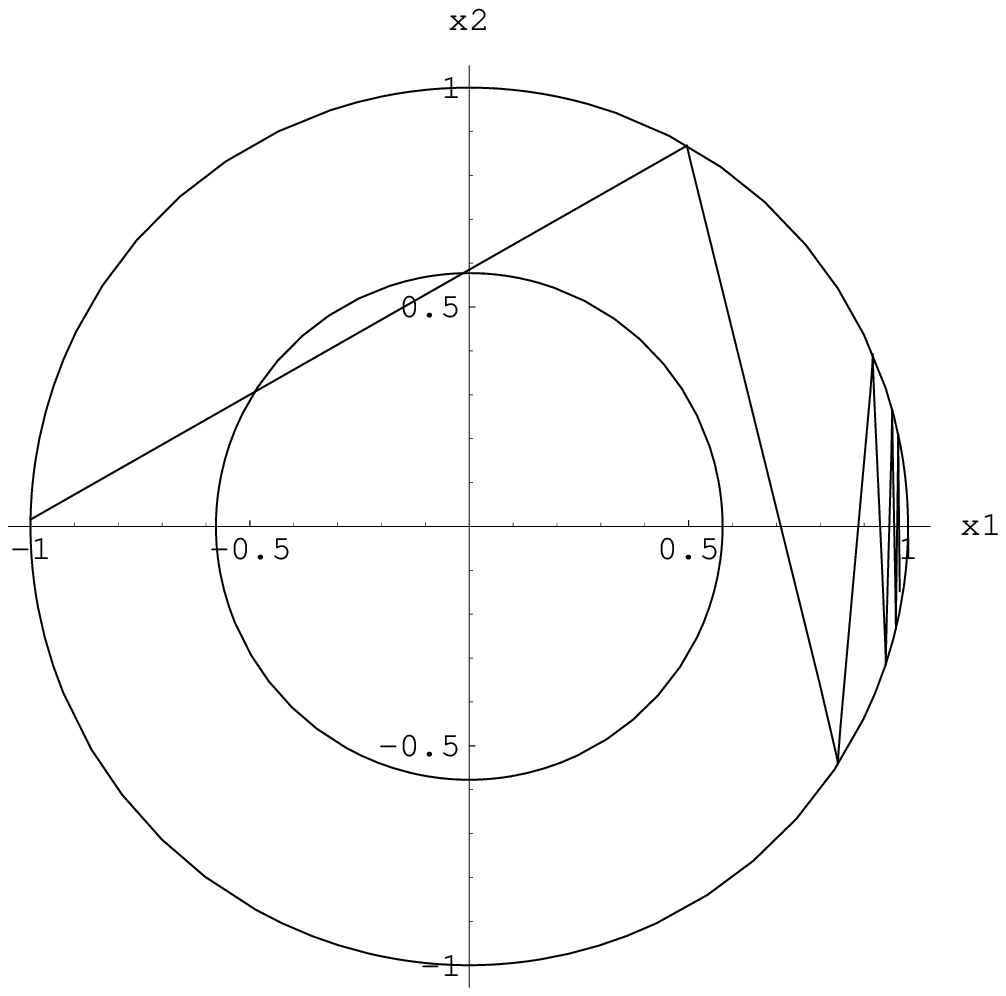,width=.23\textwidth}\hspace{1cm}
\epsfig{file=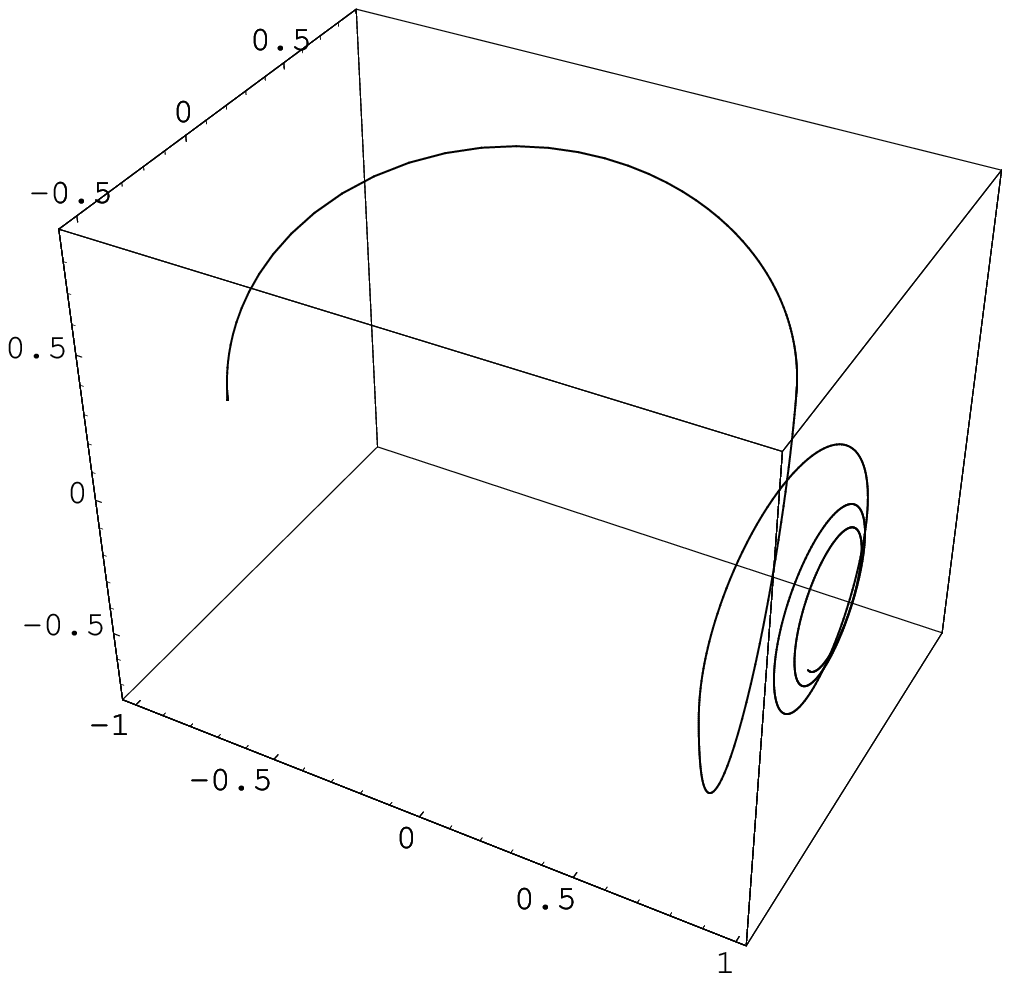,width=.23\textwidth}}
 \caption{\it {\small Type VII$_0$. To the left are shown the projection of two
 interpolating solutions on the $(x_1,x_2)$-plane.
 To the right, the curves are shown on the 2-sphere defined by the Friedmann equation;
 the vertical axis is given by $y_1-y_2$. The fact that the curves
 do not reach the attractor is due to the finite computation time.}}
\label{sixnill}
\end{figure}

For the sake of illustration, $y_1\pm y_2$ can be used as a
variable, such that any solution will be given by points or curves
on a 2-sphere (the upper half in case of the plus sign, since the
$y$'s are positive). Interestingly, the dilaton couplings are such
that most of the critical points from the previous section do not
exist. In fact, for type VI$_0$ we are just left with the equator
solutions and for type VII$_0$ we have the equator and an infinite
set of proper solutions
\begin{eqnarray}  \nonumber
&\qquad\ x^2=1\,, \qquad\qquad\ y_1=y_2=0\qquad\ \quad{\rm type}\quad {\rm VI}_0\,,\\
&\left.\begin{array}{cl}x^2=1\,,& \qquad y_1=y_2=0\\
(x_1,x_2)=(1,0)\,,& \qquad y_1=y_2
\end{array}\right\}\qquad{\rm type}\quad {\rm VII}_0\,.
\end{eqnarray}
By studying the derivatives of the coordinates, it can be shown
that the following points are attractors
\begin{alignat}{3}  \nonumber
(x_1,x_2)&=(1,0) & & \qquad y_1=y_2=0 & & \qquad{\rm type}\quad {\rm VI}_0\,,\\
(x_1,x_2)&=(1,0) & & \qquad y_1=y_2 & & \qquad{\rm type}\quad {\rm
VII}_0\,.
\end{alignat}

Thus, the latter is not unique since the $y_i$-values are
arbitrary. The solution corresponds to \eqref{xsolution2}, which
is possible because of the convex combination:
$\vec{\alpha}_3=(\vec{\alpha}_1+\vec{\alpha}_2)/2$. In both cases
any interpolating solution will end in the point $(1,0,0)$ on the
2-sphere. In case VII$_0$, the $y$-value will be determined by the
initial conditions. The sign of $\dot{x}_1$ is always positive.
When projected on the $(x_1,x_2)$-plane, any curve will therefore
move from left to right. 

A stability analysis  leads to the result that only the part of
the equator with $x_1<-1/7$ is repelling. Thus, any interpolating
solution can start on this part and will end in $(1,0,0)$.

\begin{figure}[ht]
\centerline{\epsfig{file=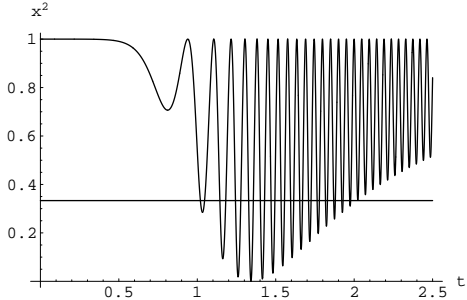,width=.4\textwidth}}
 \caption{\it {\small Type VII$_0$. An example of a $(t,x^2)$-plot with $n<1$.}}
 \label{efold}
\end{figure}

A couple of typical curves for interpolating solutions with
different initial conditions are depicted on figure \ref{sixnill}
for type VII$_0$. In this case, any curve will be
 spiralling around the 2-sphere towards the attractor. The projection on the $(x_1,x_2)$-plane
produces a curve which bounces off the boundary of the unit
circle. The inner disc, corresponding to $x^2<1/3$, yields phases
of accelerate expansion. Depending on the initial conditions, the
number of such phases can be as high or low as we want. For the
two cases on figure \ref{sixnill}, the numbers are 16 and 1,
respectively. Even with a lot accelerating phases, the number of
e-foldings is of order 1, and therefore these models are not well
suited for inflation. The numerical solutions use $t=\ln (a)$ as
time parameter. The number of e-foldings is
\begin{align}
n=\ln \Big(\frac{a(\tau_2)}{a(\tau_1)}\Big)=\ln
\Big(\frac{e^{t_2}}{e^{t_1}}\Big)=\Delta t\,,
\end{align}
and its order of magnitude can easily be read off from a
$(t,x^2)$-plot as the sum of the $t$-intervals where $x^2<1/3$. An
example is given in figure \ref{efold}.

\begin{figure}[ht!]
\centerline{\epsfig{file=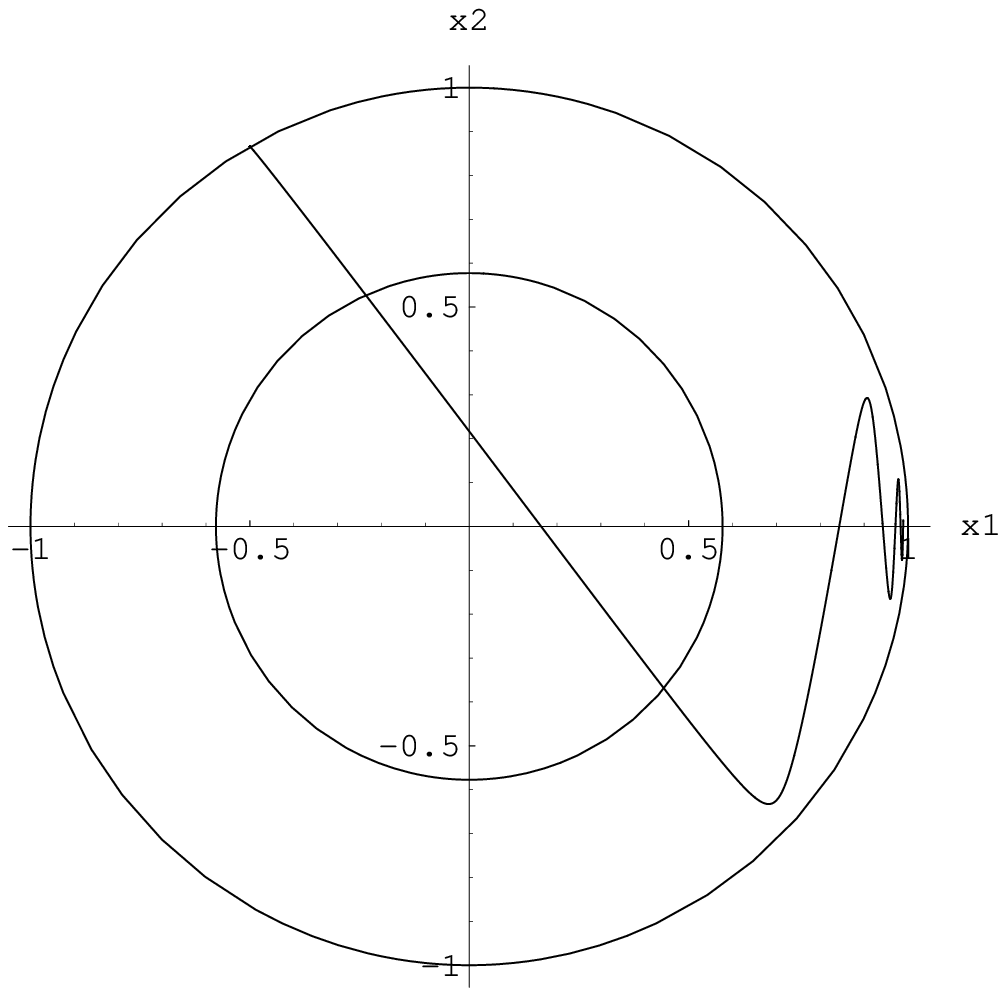,width=.23\textwidth}\hspace{1cm}
\epsfig{file=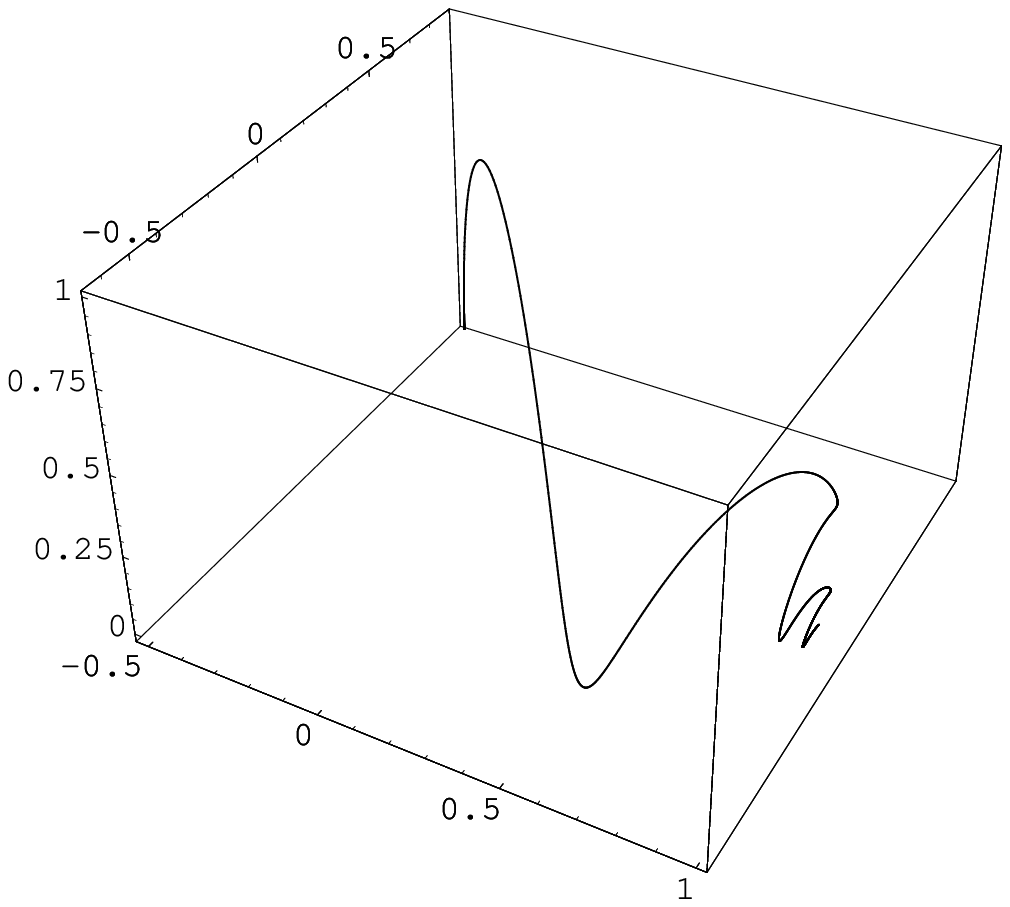,width=.23\textwidth}}
\vspace{0.5cm}
\centerline{\epsfig{file=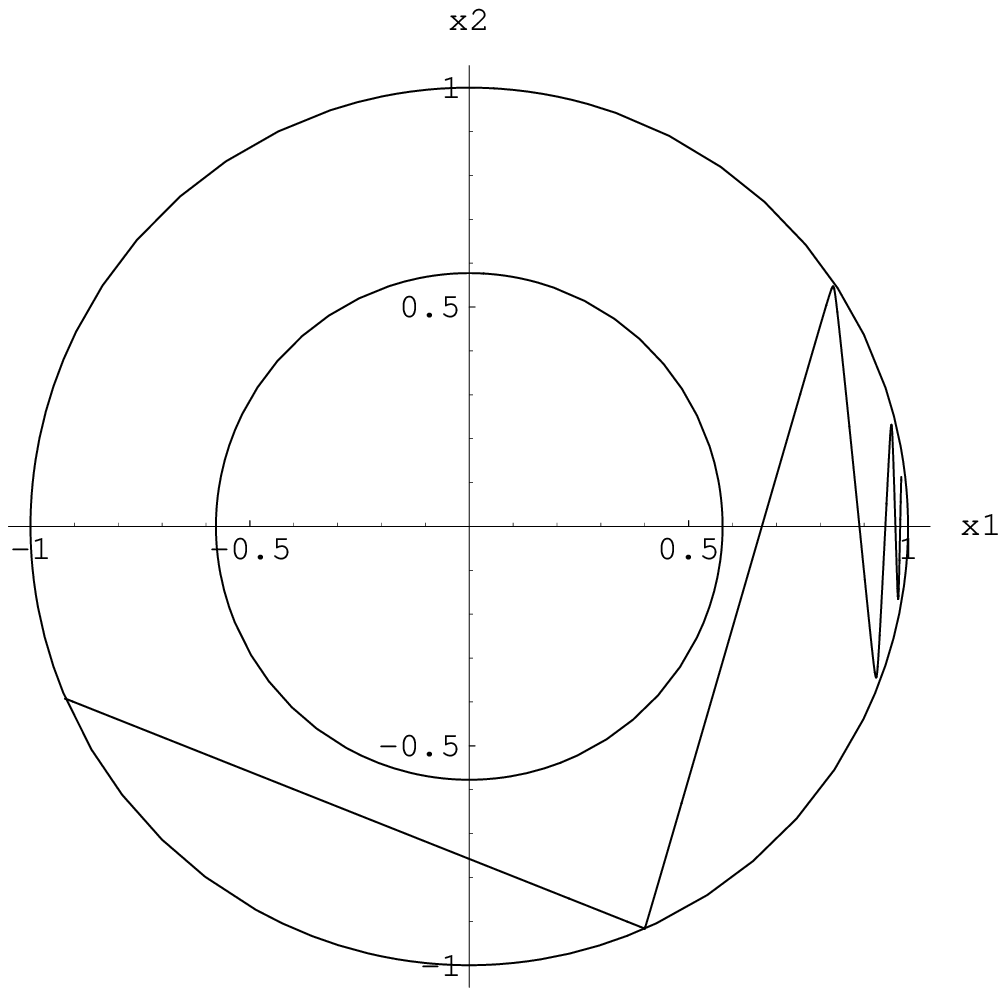,width=.23\textwidth}\hspace{1cm}
\epsfig{file=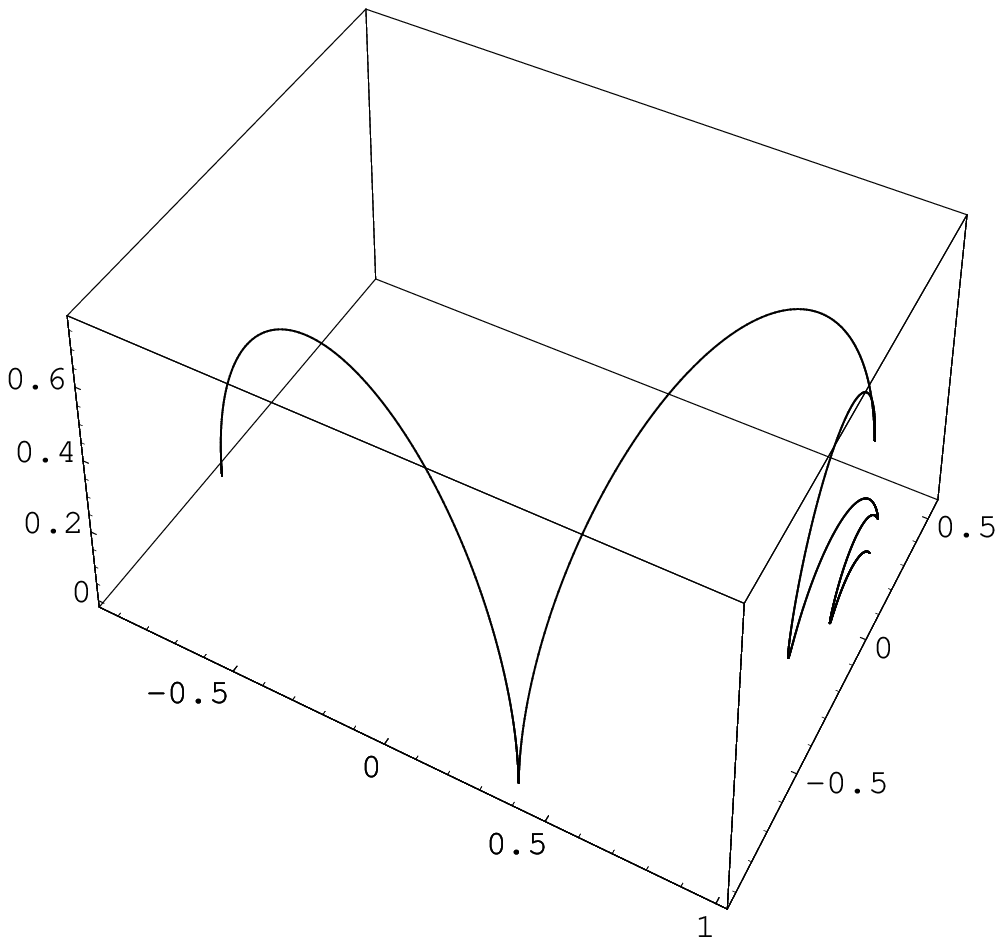,width=.23\textwidth}}
 \caption{\it {\small Type VI$_0$. To the left are shown the projection of two
 interpolating solutions on the $(x_1,x_2)$-plane.
 To the right, the curves are shown on the 2-sphere defined by the Friedmann equation;
 the vertical axis is given by $y_1+y_2$.}}
 \label{sevennill}
\end{figure}

For type VI$_0$, the situation is slightly different, since
$y_1+y_2$ is always positive; this confines the curves to the
upper half of the 2-sphere. As seen on figure \ref{sevennill}, the
curves will still move towards the attractor in an oscillatory
manner, but now without crossing the equator (though they can get
arbitrarily close). For this case, there can only be one or no
phase of accelerate expansion.

The interpolating solutions above correspond to reductions of
exotic S2-branes in five dimensions, or equivalently, exotic
S$(D-3)$-branes in $D$ dimensions. However, the solutions were
found numerically,  and we have not been able to obtain exact
expressions for these exotic S-branes.

\section{Discussion}

In this paper we have considered cosmological models for an
arbitrary number of scalars with an arbitrary multi-exponential
potential. Using a special set of variables, the equations were
written as an autonomous dynamical system, and this allowed us to
determine the critical points in complete generality. We found
that the nature of these critical points depends strongly on the
rank $R$ of the matrix $\alpha_{iI}$. The rank also determines the
number of decoupled scalars.

In the case $R=m$, both the proper and non-proper critical points
are power-law solutions, and there are no de Sitter solutions. In
the $R<m$ case the opposite behaviour was found. Proper power-law
solutions are only possible in special cases, where the
$\vec{\alpha}_i$-vectors are linearly dependent in a specific way,
but the de Sitter solutions are very generic. For the non-proper
solutions this depends on whether the ``truncated'' potential has
$R=m$ or $R<m$. We also found a new property of these systems,
namely the possibility of proper critical points which are not
unique. A special case was realised in section 5, where we have an
infinite set of these.

We would like to emphasise that the non-proper critical points are
as important as the proper solutions in understanding the
interpolating solutions, even though they have not been considered
often in the literature. In this respect, using the techniques of
autonomous systems is more fruitful than simply looking for
power-law solutions to the equations of motion.

It should be pointed out that our solutions generically have
run-away behaviour of the scalars. The only exceptions are the de
Sitter critical points, since these correspond to an extremum of
the potential and accordingly stabilise the values of the scalars.
This is important in the context of spontaneous decompactification
\cite{Giddings:2004vr} or stabilisation of dilaton and volume
moduli \cite{Kachru:2003aw}

In section 4 and 5 we gave several examples of double and triple
exponential potentials. We presented the critical points and
illustrated the interpolating solutions using numerical
calculations. In particular, we found examples with an arbitrarily
high number of phases of accelerate expansion. However, the number
of e-foldings turns out to be of order one, so these models do not
seem to be relevant for inflation. On the other hand, they might
be relevant for present-day acceleration and more specifically for
solving the cosmic coincidence problem.

The numerical solutions found in section 5 for the systems
obtained from reductions over group manifolds of type VI$_0$ and
type VII$_0$ correspond to the reduction of exotic S2-branes in
five dimensions. The two solutions belong to a set of three
different solutions, which can be obtained via twisted circle
reductions. The third solution can be obtained from a reduction
over the type II group manifold and corresponds to the reduction
of a fluxless S2-brane. The existence of three classes of S-branes
is similar to the cases of 7-branes in ten dimensions
\cite{Bergshoeff:2002mb} and non-extremal D-instantons
\cite{Bergshoeff:2004fq} and is reminiscent of the global
$SL(2,\mathbb{R})$-symmetry of the higher dimensional theory. It
would be interesting to see whether it would be possible to find
exact solutions for the exotic S-branes.

Recently, an elegant framework for arbitrary potentials has been
developed, where the solutions correspond to geodesics in an
augmented target space \cite{Townsend:2004zp}. One of the key
ingredients is the importance of systems whose late-time behaviour
is governed by single exponential potentials
\cite{Wohlfarth:2003kw}. In our analysis these solutions asymptote
to the special class of non-proper critical points where all $y$'s
but one vanish. However, we have shown that multi-exponential
potentials have solutions, where the asymptotics can not governed
by a single term in the potential. Specific examples are given by
the cases of assisted inflation \cite{Liddle:1998jc} and
generalized assisted inflation \cite{Copeland:1999cs} where each
term in the potential contributes.

We would like to comment on some possible extensions of this work.
First of all, we have only considered flat universes, and it is
certainly possible to extend to the spatially curved cases.
Secondly, we could also add matter,in the form of a barotropic
fluid. This has been done in the case of assisted inflation in
\cite{Coley:1999mj}. In the same spirit we hope to extend this to
the most general exponential potential and report on it in a
future publication. Thirdly, we could consider non-flat scalar
manifolds. Finally, we could consider other specific numerical
examples with other values of $m$ and $N$ and special dilaton
couplings which arise from dimensional reductions of
string/M-theory.

\section*{Acknowledgements}

We thank Eric Bergshoeff, Martijn Eenink, Diederik Roest and Ulf
Gran for interesting discussions. The work of Thomas Van Riet is
part of the research programme of the ``Stichting voor
Fundamenteel Onderzoek van de Materie'' (FOM). This work is
supported in part by the European Community's Human Potential
Programme under contract HPRN-CT-2000-00131 Quantum Spacetime, in
which A.C., M.N.~and T.V.R. are associated to Utrecht University.

\bibliography{accelerating}
\bibliographystyle{utphysmodb}

\end{document}